\pdfoutput=1
\documentclass[english,preprint,showpacs,preprintnumbers,amsmath,amssymb]{revtex4}
\usepackage[T1]{fontenc}
\usepackage[latin9]{inputenc}
\usepackage{graphicx}

\makeatletter

\providecommand{\tabularnewline}{\\}

\@ifundefined{textcolor}{}
{%
 \definecolor{BLACK}{gray}{0}
 \definecolor{WHITE}{gray}{1}
 \definecolor{RED}{rgb}{1,0,0}
 \definecolor{GREEN}{rgb}{0,1,0}
 \definecolor{BLUE}{rgb}{0,0,1}
 \definecolor{CYAN}{cmyk}{1,0,0,0}
 \definecolor{MAGENTA}{cmyk}{0,1,0,0}
 \definecolor{YELLOW}{cmyk}{0,0,1,0}
 }

\usepackage{bm}

\makeatother

\usepackage{babel}

\begin{document}

\title{  New ultrahigh pressure phases of H$_2$O ice predicted using an adaptive genetic algorithm }

\author{ Min Ji$^{1}$, Koichiro Umemoto$^{2}$, Cai-Zhuang Wang$^{1}$,
Kai-Ming Ho$^{1}$, and Renata M. Wentzcovitch$^{3}$}

\affiliation{ $^{1}$Ames Laboratory, US DOE and Department of Physics and Astronomy,
Iowa State University, Ames, Iowa 50011, USA \\
 $^{2}$Department of Geology and Geophysics, University of Minnesota,
Minneapolis, MN 55455, USA\\
 $^{3}$Minnesota Supercomputing Institute and Department of Chemical
Engineering and Materials Science, University of Minnesota, Minneapolis,
MN 55455, USA }

\date{\today}
\begin{abstract}
We propose three new phases of H$_{2}$O under ultrahigh pressure.
Our structural search was performed using an adaptive genetic algorithm which allows an extensive exploration of crystal structure. The new sequence
of pressure-induced transitions beyond ice X at 0 K should be ice
X$\to Pbcm\to Pbca\to Pmc2_{1}\to P2_{1}\to P2_{1}/c$ phases. Across
the $Pmc2_{1}$-$P2_{1}$ transition, the coordination number of oxygen
increases from 4 to 5 with a significant increase of density. All
stable crystalline phases have nonmetallic band structures up to 7
TPa. 
\end{abstract}
\maketitle



H$_{2}$O ice is one of the most abundant planet forming materials
and its phase diagram is of fundamental scientific interest. It is
crucial for modeling the interiors of icy solar giants (Uranus and
Netpune), icy satellites, and also new ocean planets being discovered
right now. Up to now, sixteen crystalline phases have been identified
experimentally \cite{Petrenko,Lobban1998,Salzmann2006,Salzmann2009}.
Ice X is the highest-pressure phase among those identified experimentally.
In this ionic phase, which exists above $\approx$ 70 GPa, oxygen
atoms form a bcc lattice and hydrogen atoms are located at the midpoint
between two nearest-neighbor oxygen atoms. However, in Uranus and
Neptune, pressure at their core-envelope boundary is estimated to
be as high as 0.8 TPa \cite{Guillot2004}. Very recently, Neptune-sized
icy exoplanets have been discovered \cite{Butler2004,Lissauer2011}.
Transitions beyond ice X can occur under these high pressure conditions
and they are essential for modeling the interiors of these planets.
Previously, Benoit et al. \cite{Benoit1996} and Caracas \cite{Caracas2008}
predicted a phase with $Pbcm$ symmetry as the next high-pressure
phase after ice X (up to 0.3 TPa). More recently, Militzer and Wilson
proposed transitions from the $Pbcm$ to $Pbca$ and $Cmcm$ phases
at 0.76 and 1.55 TPa, respectively \cite{Militzer2010}. Interestingly,
they predicted the $Cmcm$ phase to be metallic, an important property
for understanding the origin of magnetic fields in the giants. While
these studies surely predict the existence of new phases, their searches
using MD and phonon instability calculations explored limited regions
of phase space producing structure relatively close to that of ice
X. In this Letter, we report new structures of solid H$_{2}$O in
the TPa pressure regime. They were found using a novel, general, and
efficient structural search algorithm, the adaptive genetic algorithm.


The search for the lowest-enthalpy structures of ultrahigh-pressure
ice is based on the Deaven-Ho genetic algorithm (GA) scheme \cite{Deaven1995}.
This method combined with first-principles calculation has been proven
to work very efficiently \cite{Oganov2006,Trimarchi2008,Ji2010,Wu2011}.
However, first-principles calculations are computationally very demanding
for the GA scheme when there are a large number of atoms. On the other
hand, GA searches using empirical model potentials are fast but suffer
from inaccuracies which can lead the search to wrong structures. However,
most of the child structures generated in GA are actually not favorable
except in the simplest problems; many false structures have to be
tried before hitting on the correct structure. Most of the computer
time is spent in relaxing a large number of false candidate structures.
To remedy this, we introduced a new structure search technique, the
adaptive GA. In the new scheme, we employ auxiliary model potentials
to estimate the energy ordering of different competing geometries
in a preliminary stage. Parameters of the auxiliary potentials are
adaptively adjusted to reproduce first-principles results during the
course of the GA search. This adaptive approach also allows the system
to hop from one basin to another in the energy landscape leading to
efficient sampling of configuration space. 

    In our study, we found that the packing geometry, volumes(pressures) and energy ordering of ice crystal phases at high pressures (>0.5 TPa) can be described relatively well by simple Lennard-Jones (LJ) potentials: 
\begin{equation}
V_{LJ}=4\varepsilon\left[\left(\frac{\sigma}{r}\right)^{12}-\left(\frac{\sigma}{r}\right)^{6}\right].
\end{equation}
In the H$_{2}$O system, a total of six parameters, representing  the O-O, O-H and H-H interactions, are adaptively adjusted to explore structural phase space during the GA search. We initiate the search with random structures and carry out DFT calculations to get their energies, atomic forces
and stress tensor. The LJ potential parameters are fitted
to these structure by force-matching method\cite{Brommer}. Based
on this auxiliary potential we perform full GA search to yield new structural candidates. DFT calculations
are carried out on the new GA pool and potential parameters are adaptively adjusted again. These procedures can be done iteratively until the potential
parameters converge. The final LJ potential pool population will then be examined with full DFT relaxation. This adaptive scheme allows a very efficient sampling of crystal structures, we can easily search relatively large systems up to 12 H$_{2}$O formula units.

Details of the GA search can be found in several previous papers \cite{Oganov2006,Trimarchi2008,Ji2010,Wu2011}.
For auxiliary potential optimization we use LAMMPS code \cite{LAMMPS}
and conjugate gradient method. First principles calculations are performed
using density-functional theory (DFT) within PBE-type generalized-gradient
approximation (GGA) \cite{Perdew1996}. Vanderbilt-type pseudopotentials
\cite{Vanderbilt1990} were generated using the following valence
electron configurations: $1s^{1}$ and $2s^{2}2p^{4}$ with cutoff
radii of 0.5 and 1.4 a.u for hydrogen and oxygen, respectively. Candidate
structures obtained with the adaptive GA were then refined using a
harder oxygen pseudopotential with valence electron configuration
of $2s^{2}2p^{4}3d^{0}$ and cutoff radius of 1.0 a.u. The cutoff
energy for plane-wave expansions are 40 Ry and 120 Ry for the softer
and the harder pseudopotentials respectively. Brillouin-zone integration
was performed using the Monkhorst-Pack sampling scheme \cite{Monkhorst1976}
over \textbf{k}-point meshes of spacing $2\pi\times0.05$\AA{}$^{-1}$.
Structural relaxations were performed using variable-cell-shape molecular
dynamics \cite{Wentzcovitch1991,Wentzcovitch1993}. To test for structural
stability, phonon and vibrational density of states (VDOS) calculations
were carried out for candidate structures using density-functional-perturbation
theory \cite{Giannozzi1991,Baroni2001}. Zero-point motion (ZPM) effects
are taken into account within the quasiharmonic approximation \cite{Wallace1972}.
All first-principles calculations were performed using the Quantum-ESPRESSO
\cite{Giannozzi2009}, which has been interfaced with the GA scheme
in a fully parallel manner.

\begin{figure}
\hbox to \hsize{\hfill
\includegraphics[width=100mm]{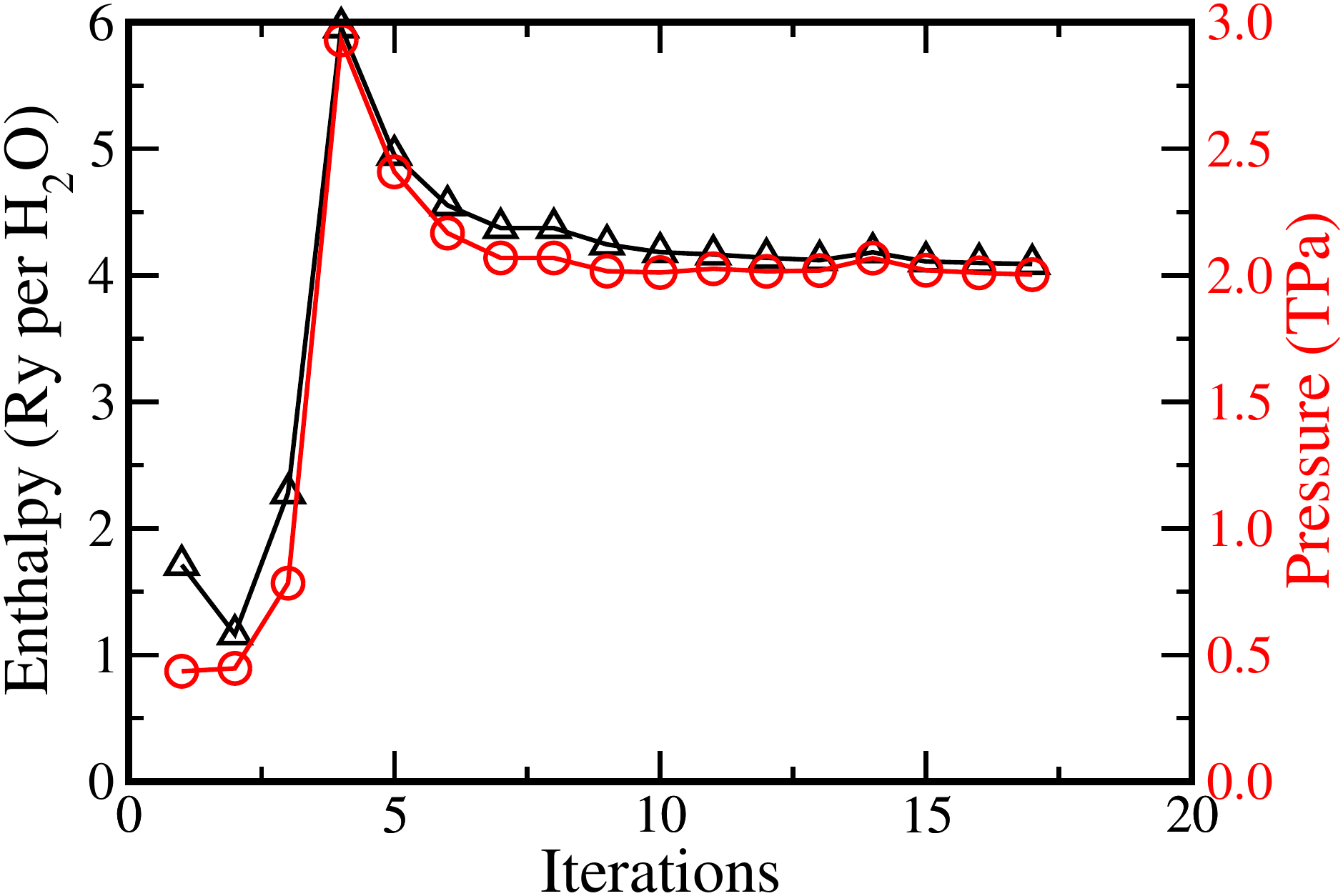}
    \hfill
 }
\caption{ Convergence of average DFT pressure and enthalpy of structures in the LJ-potential GA pool as a function of adaptive potential iterations. The DFT pressure converges to the target pressure of 2 TPa after ~10 iterations. The unit cell contains
8 H$_2$O units.}

\label{LJ} %
\end{figure}

\begin{figure}
\hbox to \hsize{\hfill
\includegraphics[width=150mm]{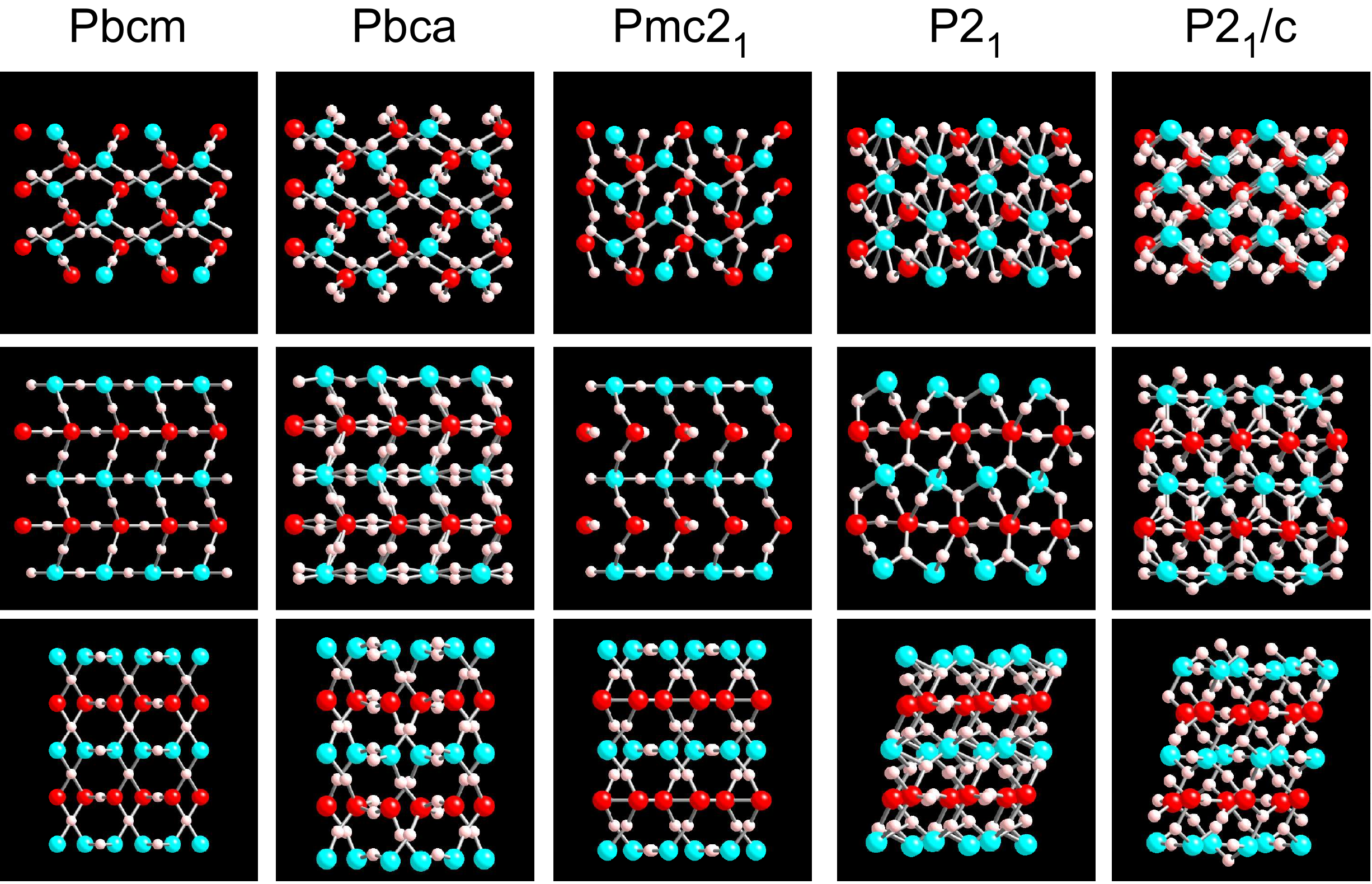}
    \hfill
 }
\caption{ Crystal structures of ultrahigh-pressure phases of ice. Blue and
red large spheres denote oxygen atoms. White small spheres denote
hydrogen atoms. }

\label{structure} %
\end{figure}

\begin{figure}
\hbox to \hsize{\hfill
\includegraphics[width=100mm]{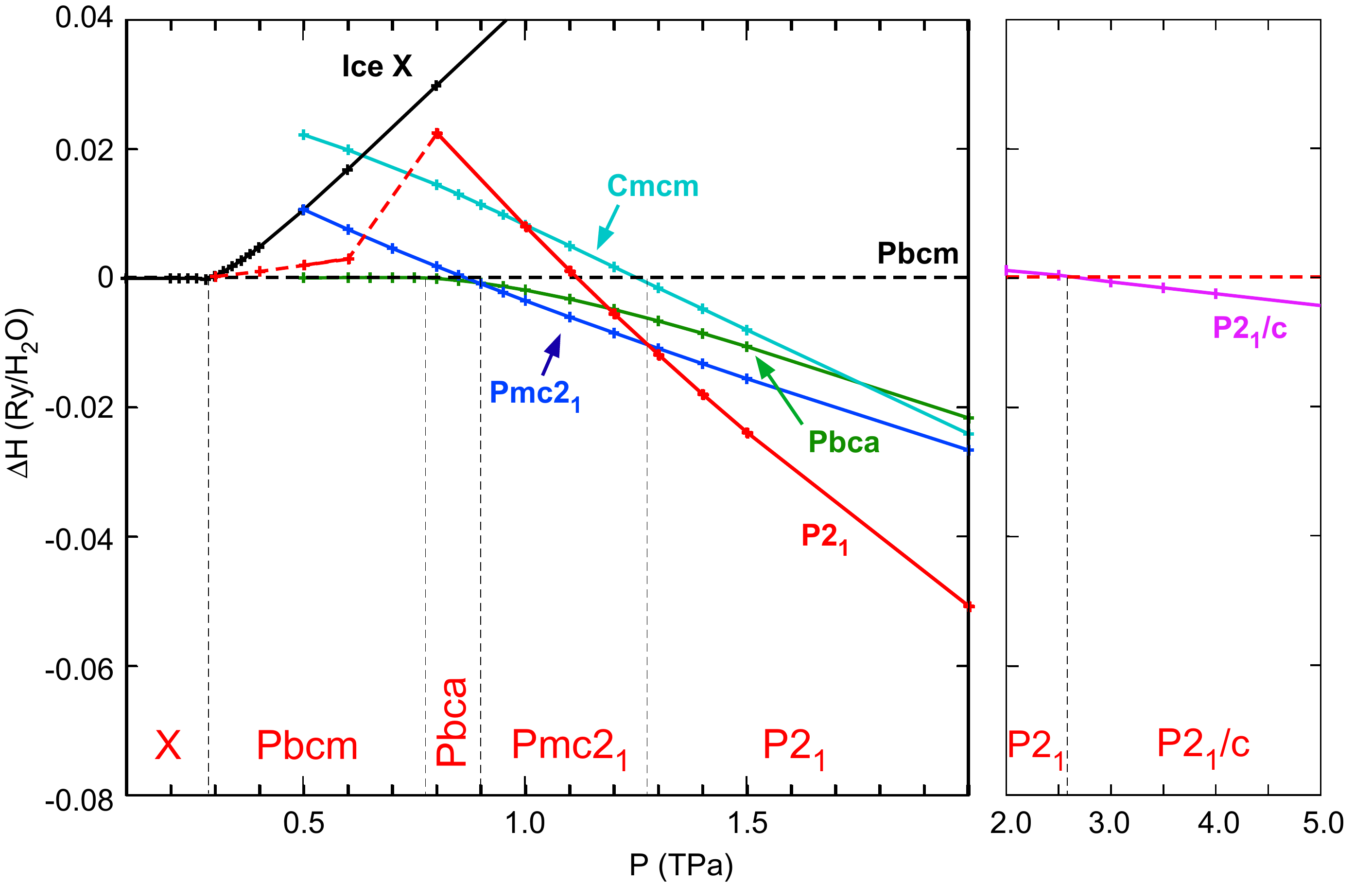}
    \hfill
 }
\caption{ Enthalpies of ultrahigh-pressure phases of ice. Dashed vertical lines
denote static transition pressures. }

\label{dH} %
\end{figure}

Fig.~\ref{LJ} shows average values of the pressures and enthalpies calculated by
first principles for structures in the adaptive GA pool. Target pressure
of this adaptive GA search is 2 TPa. Our results show a fast convergence
of the adaptive GA. After 10 iterations, LJ-potential pressures for
structures in the GA pool are almost identical to first-principles
DFT results. The adaptive GA search successfully predicts three new
structures: $Pmc2_{1}$ at 1 TPa, $P2_{1}$ at 2 TPa, and $P2_{1}/c$
phase at 3 TPa (Fig.~\ref{structure}). They consist of 4, 4, and
8 formula units, respectively. Structures with 3, 5, 6, 7, 9, 10, and 11 formula
units were not energetically competitive. $Pbcm$ and $Pbca$ phases
proposed previously were also obtained as metastable phases at these
pressures. Examination of the enthalpy differences between the different
competing phases (Fig.~\ref{dH}) shows that the sequence of pressure-induced
phase transitions beyond ice X is X$\to Pbcm\to Pbca\to Pmc2_{1}\to P2_{1}\to P2_{1}/c$.
Static transition pressures are 0.28, 0.75, 0.89, 1.28, and 2.68 TPa,
respectively. Zero-point motion strongly affects transition pressures.
For X-$Pbcm$, $Pbcm$-$Pbca$, and $Pbca$-$Pmc2_{1}$ transitions,
ZPM increases transition pressures to 0.29, 0.77, and 0.92 TPa. On
the other hand, for $Pmc2_{1}$-$P2_{1}$ and $P2_{1}$-$P2_{1}/c$
transitions, ZPM greatly decreases transition pressures to 1.14 and
1.96 TPa.

The structures of these competing phases are closely related. Under
compression, ice X transforms to the $Pbcm$
phase and then to the $Pbca$ phase by means of soft phonon related deformations \cite{Caracas2008,Militzer2010}.
In these three phases, all hydrogen atoms are located at the midpoint
between two neighboring oxygen atoms. However, during the transition
to the $Pmc2_{1}$ phase, two among the four hydrogen atoms bonded
to the O1 oxygen jump to the octahedral interstitial sites next to
two second-nearest-neighbor oxygen atoms. The arrangements of hydrogen
atoms are $Pbca$-like around O1 and $Cmcm$-like around O2. Therefore,
this phase is structurally intermediate between the $Pbca$ and $Cmcm$
phases. Its becomes metastable with respect to the $Cmcm$ phase at
$\sim$2.5 TPa. In ice X, $Pbcm$, $Pbca$, and $Pmc2_{1}$ phases,
OH$_{4}$ tetrahedra form the basic structural unit, with connectivity
varying in each phase. Across the $Pbca$ to $Pmc2_{1}$ transition,
two interpenetrating tetrahedral networks transform into a single
network. Locally, hydrogen atoms around O2 try to keep two interpenetrating
networks, while those around O1 atoms connect two networks into a
single one. Tetrahedra around the O1 atoms are severely distorted
compared to those around the O2 atoms.

The $Pmc2_{1}$ phase is dynamically stable up to 1.3 TPa. At $\sim$1.5
TPa, a zone-center soft mode appears. This soft mode induces a monoclinic
distortion giving rise to the $P2_{1}$ phase. The $P2_{1}$ space
group is a subgroup of $Pmc2_{1}$. In the $P2_1$ phase, the structural
unit is no longer OH$_{4}$. The coordination number of oxygen atoms
increases from 4 to 5. In X, $Pbcm$, $Pbca$, and $Pmc2_{1}$ phases,
phonon dispersions are divided into two groups corresponding to nearly-rigid-body
and internal O-H stretching motions of OH$_{4}$ tetrahedra, while 
this distinction is blurred in the $P2_1$ phase. As a
result of higher coordination number, increase in H-O bond-length,
and a 2.0\% volume reduction across the $Pmc2_1$-$P2_1$ transition.
Both $P2_{1}$ and $P2_{1}/c$ phases are dynamically stable at least up to 7 TPa, with
no soft modes developing under compression.

\begin{figure}
\hbox to \hsize{\hfill
\includegraphics[width=100mm]{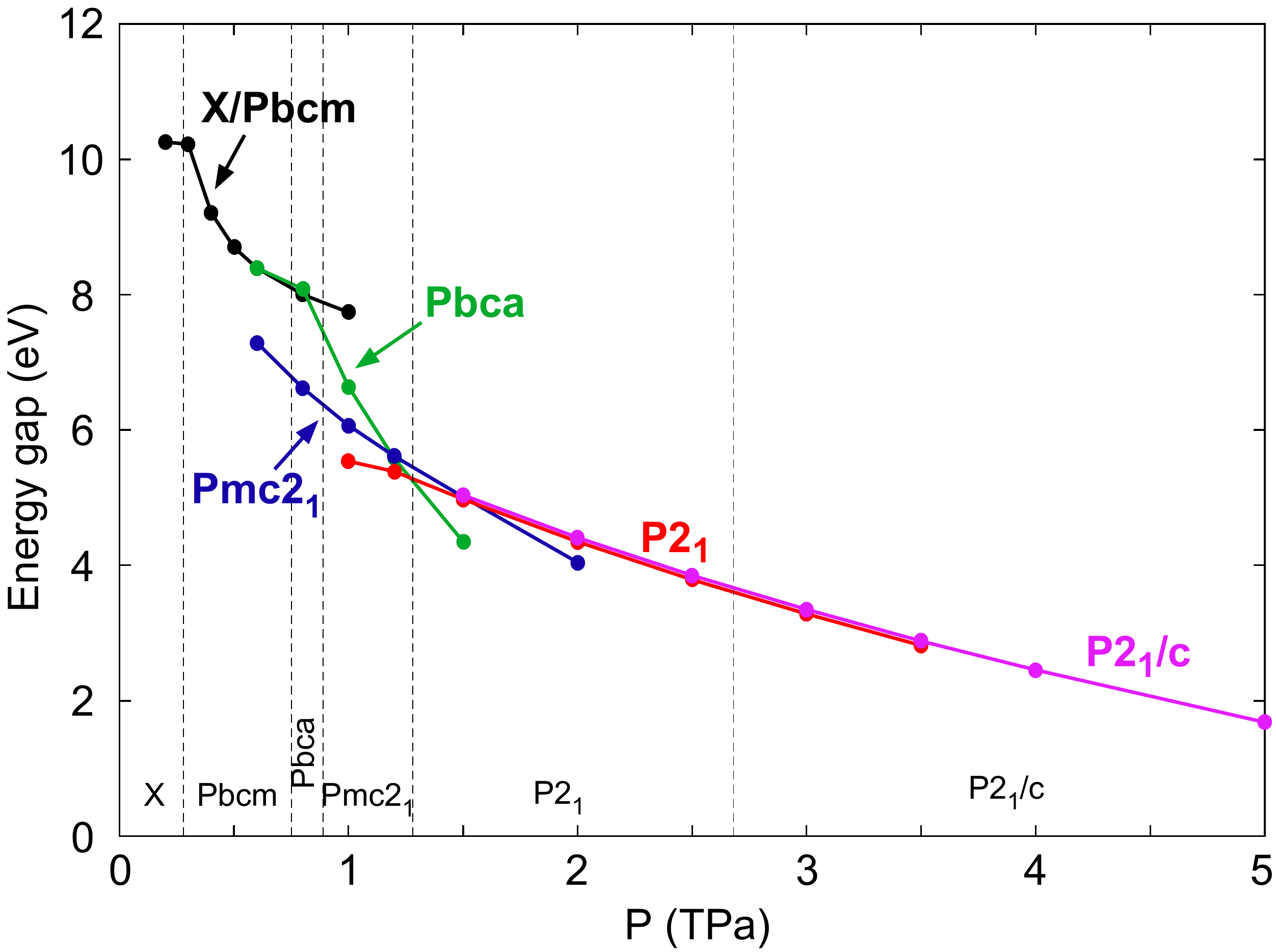}
    \hfill
 }
\caption{ Electronic band gaps of ultrahigh-pressure phases of ice. Dashed
vertical lines denote static transition pressures. }

\label{gap} %
\end{figure}

In contrast with the metallic $Cmcm$ phase predicted by \cite{Militzer2010},
all three new phases ($Pmc2_{1}$, $P2_{1}$, $P2_{1}/c$) have substantial
DFT band gaps. In the $P2_{1}$ and $P2_{1}/c$ phases, the band gap decreases
slowly under compression (Fig.~\ref{gap}), closing at $\sim$7 TPa
in the $P2_{1}/c$ phase. Although all crystalline phases are insulating,
H$_{2}$O could be a good conductor at relevant conditions for two
reasons: 1) protons highly mobile in the oxygen sublattice producing
superionic phases at ultrahigh pressures and temperatures typical
of the interiors of icy solar giants and exoplanets; 2) depending
on the temperature, thermally excited carriers also contribute to
increase the conductivity \cite{Cavazzoni1999,Chau2001,Goldman2005,Goncharov2005,Mattsson2006,Schwegler2008,French2009,Redmer2011}.
So far, only the bcc oxygen sub-lattice has been considered in the
investigation of superionic phases. Our new crystal structures provide
starting points for further investigation of conducting superionic
states of these phases at ultrahigh pressures and temperatures. They
indicate that the oxygen sublattice should prefer to have hexagonal-derived
structures beyond $\sim$0.4 TPa. This possibly implies a bcc-hcp
transition in the superionic phase.

Work at Ames Laboratory was supported by the US Department of Energy,
Basic Energy Sciences, Division of Materials Science and Engineering,
under Contract No. DE-AC02-07CH11358, including a grant of computer
time at the National Energy Research Supercomputing Centre (NERSC)
in Berkeley, CA. KU and RMW's work were supported by NSF grants EAR-0757903,
EAR-0810272, EAR-1047629, and ATM-0426757 (VLab). Computations at
the University of Minnesota were performed at the Minnesota Supercomputing
Institute and at the Laboratory for Computational Science and Engineering.

\newpage{} %
\begin{table}
\begin{tabular}{ccc}
\hline 
\multicolumn{3}{c}{$Pmc2_{1}$-type H$_{2}$O at 1 TPa}\tabularnewline
\multicolumn{2}{c}{$(a,b,c)$}  & (3.087\AA, 1.890\AA, 3.296\AA)  \\ \tabularnewline
H$_{1}$  & $4c$  & (0.25235, 0.39936, 0.34521) \tabularnewline
H$_{2}$  & $2a$  & (0, 0.72441, 0.58937) \tabularnewline
H$_{3}$  & $2b$  & (0.5, 0.02194, 0.05554) \tabularnewline
O$_{1}$  & $2a$  & (0, 0.79553, -0.00013) \tabularnewline
O$_{2}$  & $2b$  & (0.5, 0.71235, 0.29490) \tabularnewline
\hline 
\multicolumn{3}{c}{$P2_{1}$-type H$_{2}$O at 2 TPa}\tabularnewline
\multicolumn{2}{c}{$(a,b,c,\beta)$}  & (1.711\AA, 3.066\AA, 2.825\AA, 99.83$^{\circ}$)  \\ \tabularnewline
H$_{1}$  & $2a$  & (0.03146, -0.004190, 0.97579) \tabularnewline
H$_{2}$  & $2a$  & (0.17734, 0.60082, 0.33256) \tabularnewline
H$_{3}$  & $2a$  & (0.25493, 0.38266, 0.73104) \tabularnewline
H$_{4}$  & $2a$  & (0.56268, 0.74399, 0.68952) \tabularnewline
O$_{1}$  & $2a$  & (0.82524, 0.52141, 0.52159) \tabularnewline
O$_{2}$  & $2a$  & (0.34462, 0.75497, 0.01421) \tabularnewline
\hline 
\multicolumn{3}{c}{$P2_{1}/c$-type H$_{2}$O at 3 TPa}\tabularnewline
\multicolumn{2}{c}{$(a,b,c,\beta)$}  & (2.921\AA, 2.890\AA, 3.338\AA, 117.86$^{\circ}$)  \\ \tabularnewline
H$_{1}$  & $4e$  & (-0.49599, 0.19043, 0.44690) \tabularnewline
H$_{2}$  & $4e$  & (-0.14264, 0.13088, -0.07153) \tabularnewline
H$_{3}$  & $4e$  & (-0.24986, -0.50964, -0.27537) \tabularnewline
H$_{4}$  & $4e$  & (0.21275, 0.37363, -0.03277) \tabularnewline
O$_{1}$  & $4e$  & (0.06825, -0.36013, -0.15960) \tabularnewline
O$_{2}$  & $4e$  & (-0.42335, -0.37490, 0.33745) \tabularnewline
\hline
\end{tabular}\caption{Structural parameters of $Pmc2_{1}$-, $P2_{1}$-, and $P2_{1}/c$-type
H$_{2}$O.}

\label{struct_param} %
\end{table}

\newpage{} %

\end{document}